\documentclass[aip,jcp,reprint]{revtex4-1}

\usepackage{bm}
\usepackage{amsmath}
\usepackage{amssymb}
\usepackage{graphicx}


\begin{document}

\title{Low-temperature hopping dynamics with energy disorder: Renormalization
group approach}

\author{Kirill A. Velizhanin}
\email{kirill@lanl.gov}
\address{Theoretical Division, Los Alamos National Laboratory, Los Alamos, NM 87545}

\author{Andrei Piryatinski}
\address{Theoretical Division, Los Alamos National Laboratory, Los Alamos, NM 87545}

\author{Vladimir Y. Chernyak}
\email{chernyak@chem.wayne.edu}
\address{Theoretical Division, Los Alamos National Laboratory, Los Alamos, NM 87545}
\address{Department of Chemistry, Wayne State University, 5101 Cass Ave, Detroit, MI 48202}

\begin{abstract}
We formulate a real-space renormalization group (RG) approach for efficient numerical analysis of the low-temperature hopping dynamics in energy-disordered lattices. The approach explicitly relies on the time-scale separation of the trapping/escape dynamics. This time-scale separation allows to treat the hopping dynamics as a hierarchical process, RG step being a transformation between the levels of the hierarchy. We apply the proposed RG approach to analyze hopping dynamics in one- and two-dimensional lattices with varying degrees of energy disorder, and find the approach to be accurate at low temperatures and computationally much faster than the brute-force direct diagonalization. Applicability criteria of the proposed approach with respect to the time-scale separation and the maximum number of hierarchy levels are formulated. RG flows of energy distribution and pre-exponential factors of the Miller-Abrahams model are analyzed. 
\end{abstract}

\maketitle

\section{Introduction}

Understanding the dynamics of diffusion and transport in systems and materials
with disorder is of paramount importance in multiple branches of modern
science and engineering. The areas where the disorder-related phenomena
are ubiquitous include biology,\cite{Dill2012-1042} condensed matter
physics,\cite{Levi2011-1541} polymer\cite{Lee2006-65,Nikolai2012-882} and nano science.\cite{Hermosa2013-1} Carriers of various nature can be involved into transport phenomena in different systems: excitons in molecular or semiconductor quantum dot aggregates and polymer chains/films; charge carriers
(electrons or holes) in disordered semiconductors, phonons and photons in disordered heat conductors and photonic crystals, respectively. 

The nature of disorder can also vary a great deal from system to system.
One of the most important sources of disorder is the energy dispersion
-- the dynamics of a system occurs in a configuration space with
a rugged landscape so that potential barriers have to be overcome, e.g., by an exciton to migrate within a polymer chain. Specifically,
it is well documented that in amorphous conjugated polymers, excitons
can be localized in short ordered segments of a polymer chain, where the energy of the exciton depends on the length of the segment giving
rise to the energy disorder. An inevitable consequence is that the
exciton, generated at an arbitrary site in the excitonic density of
states distribution, will migrate (spectral diffusion) and, ultimately,
tend to relax towards low-energy states.\cite{Arkhipov2004-205205,Anh2007-43,Athanasopoulos2008-11532,Kohler2011-4003,Hwang2011-610}
It has to be noted that in amorphous films of conjugated polymers
the energy disorder can be strong in a sense that the site-to-site
energy variation can significantly exceed $k_{B}T$ at room temperature.\cite{Meskers2001-9139,Athanasopoulos2008-11532}
Possible mechanisms of exciton-exciton coupling between polymer segments
include F\"orster resonance energy transfer (due to dipole-dipole Coulomb
interaction),\cite{Forster1948-55} as well as Dexter energy transfer (due
to the direct overlap of electronic wavefunctions).\cite{Dexter1953-836}

Exciton transport in aggregates of semiconductor quantum dots has received much attention lately because of multiple
promising applications in photovoltaics, electronics and information
processing. Within such aggregates, the energy of lowest exciton varies between quantum dots due to the quantum mechanical
confinement of carriers in quantum dots with size distribution. \cite{Mohanan2005-397,Gaponik2012-8} In these systems,
photoinduced excitons have been demonstrated to efficiently migrate
within an aggregate.\cite{Crooker2002-18,Reitinger2011-710,Belyakov2011-365}
One of the possibilities to make quantum dots ``talk'' to each other,
i.e., to make an exciton migrate, is via aerogels - a special type
of aggregates - which are of great promise for photovoltaic and thermoelectric
applications.\cite{Ganguly2011-8800,DeFritas2012-15180}

Diffusion and transport phenomena in the presence of disorder are often characterized
by the absence of well-defined time, size or energy scales. The renormalization
group (RG) approach is one of the most suitable methods to study dynamics under
such conditions. The RG has been extensively applied before to study the
diffusion in energy-disordered systems, however mostly for the case of weak disorder.\cite{Deem1994-911,Bouchaud1990-127}
Transport in the case of strong disorder, i.e., where the energy dispersion significantly exceeds
the temperature (in energy units) of the environment, has been studied
to a much lesser extent mostly due to the lack of appropriate methods.\cite{Bouchaud1990-127}
In this paper, we introduce the numerical real-space RG method capable of studying hopping diffusion in the presence of a strong energy disorder.

The rest of the paper is organized as follows. The notation and general
theory is introduced in Section~\ref{sec:general}. The RG transformation is described in Section~\ref{sec:cgraining}. General numerical
results and the discussion of the RG flow are discussed in Secs.~\ref{sec:num}
and \ref{sec:flow}, respectively. Section~\ref{sec:conclusion}
concludes.

\section{Hopping on graph\label{sec:general}}

\begin{figure*}
\includegraphics[width=4.5in]{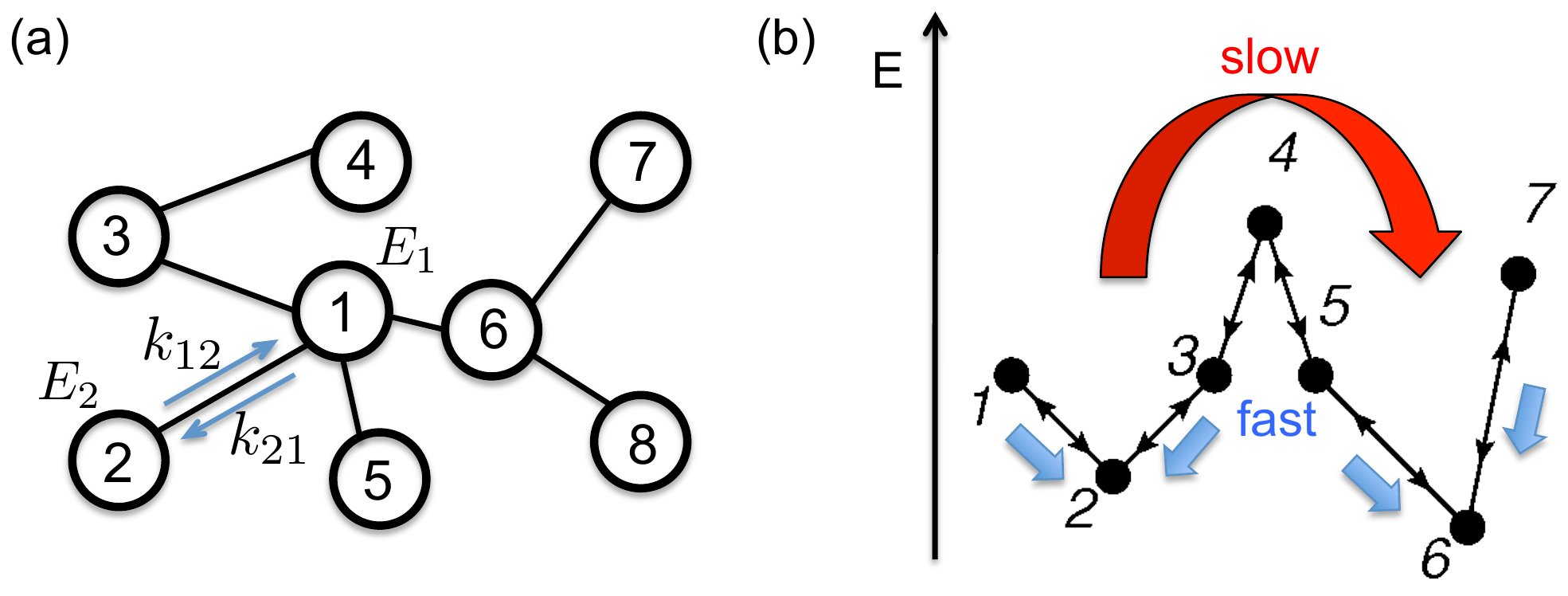}
\caption{\label{fig:graph}
(a) Schematic of a graph for hopping dynamics.
Rate constants are shown for an edge connecting nodes $1$ and $2$.
(b) Schematic of hopping dynamics on the graph with two local energy
minima. At low temperature, the relaxation dynamics within each minimum
is expected to be mush faster than the subsequent population transfer
\emph{between} the minima.}
\end{figure*}

We describe the hopping dynamics of a system with energy disorder
as Markovian hopping between nodes of a graph, where the hopping probability is defined by rate constants assigned to edges of the graph
[see Fig.~\ref{fig:graph}(a) for graph schematics].%
\footnote{This corresponds to the case where the configuration space of the
system is discrete (because the set of nodes of the graph is countable).
In Sec.~\ref{sec:conclusion} we briefly discuss how the proposed
RG method can also be applied in the continuous case.%
}
%============================================================ 
%============================================================ 
Physically, each node of this graph can represent an ordered (i.e., unbent) segment of a polymer chain
with kinks, so that at each moment of time an exciton is localized within a single segment. The time-dependent population is assigned to each node of the graph, representing the probability to find an exciton within the corresponding segment at a specific moment of time.  The length dispersion of segments results in a variation of exciton energy, thus introducing the energy disorder. Accordingly, each node of the graph is parametrized with energy. 

The probability for an exciton to hop from one node to another is specified by two hopping rate constants (forward and backward), which are assigned to each edge of the graph, as illustrated
in Fig.~\ref{fig:graph}(a) for nodes $1$ and $2$. An important
characteristic of the so formed graph is its connectivity --  the average number of edges connected
to a node. In principle, a hopping rate constant can be non-zero between
any two nodes of the graph, regardless of how distant they are from each other.
A physical example of this is a non-vanishing exciton-exciton
coupling between two very distant segments of a polymer
chain, or between distant quantum dots within an aggregate. However, hopping rate constants typically decay rapidly with distance, $R$, e.g., as $\sim R_{0}^{6}/R^{6}$ in the case of the dipole-dipole
(F\"orster) or $\sim e^{-R/R_{0}}$ in the case of the exchange-type
(Dexter) interaction, where $R_{0}$ is a characteristic range
of interaction. Henceforth, we allow edges to link only nearby nodes,
thus restricting the graph connectivity to be finite even in the thermodynamic
limit, where the number of nodes within the graph grows to infinity.

The hopping dynamics on the so introduced graph is governed by the
following kinetic equation \cite{vanKampen1992}
\begin{equation}
\partial_{t}|p(t)\rangle=\hat{W}|p(t)\rangle,\label{eq:mastereq}
\end{equation}
where $|p(t)\rangle$ is the vector of time-dependent node populations. We denote the projection of an arbitrary vector $|a\rangle$ onto node $i$ via $(i | a\rangle$, so that, e.g., $p_{i}(t)=(i|p(t)\rangle,~i=1,...,N$ ($N$ is the total number
of nodes in the graph) is the population of node $i$ at time
$t$. The rate matrix is constructed as follows.\cite{vanKampen1992}
Its off-diagonal elements read as
\begin{equation}
W_{ij}=(i|\hat{W}|j)=k_{ij},\, i\neq j\label{eq:ratec}
\end{equation}
where $k_{ij}$ is the rate constant for hopping from node $j$ to
node $i$ [see Fig.~\ref{fig:graph}(a)]. The diagonal matrix elements are not independent and are determined from the population conservation condition, which requires the sum of elements in any column of $\hat{W}$ to vanish identically, yielding
\begin{equation}
W_{ii}+\sum_{j\neq i}W_{ji}=0.\label{eq:conserv}
\end{equation}
Therefore, Eq.~(\ref{eq:mastereq}) is the gain-loss equation for
populations of nodes.

An additional constraint imposed on the rate constants is the detailed balance
\begin{equation}
\frac{k_{ij}}{k_{ji}}=e^{-\beta(E_{i}-E_{j})},\label{eq:detbal}
\end{equation}
where $\beta=1/k_{B}T=1/T$ ($k_{B}=1$ throughout the paper). The energy
of node $i$ is denoted by $E_{i}$. Eq.~(\ref{eq:detbal}) guarantees
that state $(i|p_{eq}\rangle\propto e^{-\beta E_{i}}$ is the
equilibrium state of the system, i.e., it is the steady state solution of Eq.~(\ref{eq:mastereq}), $\partial_{t}|p(t)\rangle=0$,
with no currents.

Eq.~(\ref{eq:mastereq}) can be formally solved as $|p(t)\rangle=e^{\hat{W}t}|p_{0}\rangle$,
where $|p_{0}\rangle=|p(t=0)\rangle$ is the vector of initial populations.
The rate matrix, satisfying Eqs.~(\ref{eq:conserv}) and (\ref{eq:detbal}), can be diagonalized as $\hat{W}|n\rangle=-\lambda_{n}|n\rangle$,\cite{vanKampen1992}
yielding %
\footnote{A generic non-symmetric matrix is not necessary diagonalizable,
but $\hat{W}$ is due to the detailed balance and population conservation constraints imposed.\cite{vanKampen1992}} 
\begin{equation}
|p(t)\rangle=\sum^{N-1}_{n=0}|n\rangle\langle n|p_{0}\rangle e^{-\lambda_{n}t}.\label{eq:meq_eigen}
\end{equation}
Here, we note that since $\hat{W}$ is not symmetric, both right and left, i.e., $\langle n|\hat{W}=-\langle n|\lambda_n$, eigenvectors have to be found. 
Because of the detailed balance, all the eigenvalues are strictly non-negative and the lowest eigenvalue is always zero, $\lambda_0\equiv 0$,
corresponding to the equilibrium. The normalized right and left eigenvectors corresponding to this equilibrium mode, $n=0$, are $|0\rangle=Z^{-1}\sum_i |i)e^{-\beta E_i}$ and $\langle 0|=\sum_i (i|$, respectively. The partition function is given by $Z=\sum_i e^{-\beta E_i}$.

Eq.~(\ref{eq:meq_eigen}) completely solves the problem defined
by Eqs.~(\ref{eq:mastereq})-(\ref{eq:detbal}) , since any observable which depends on
node populations at an arbitrary moment of time can now be expressed
via Eq.~(\ref{eq:meq_eigen}). For example, the average energy of an exciton at time $t$ is given by
\begin{equation}
\langle E(t)\rangle\equiv\frac{\langle E|p(t)\rangle}{\sum_i (i|p(t)\rangle}=\frac{\sum_{n}\langle E|n\rangle\langle n|p_{0}\rangle e^{-\lambda_{n}t}}{\sum_i (i|p_{0}\rangle},\label{eq:en_av}
\end{equation}
where $\langle E|$ is the vector of node energies, i.e., $\langle E|i)=E_i$. Deriving Eq.~(\ref{eq:en_av}), we explicitly
used the population conservation requirement.

\section{Coarse graining\label{sec:cgraining}}

Eq.~(\ref{eq:meq_eigen}) is an exact solution of the problem of hopping dynamics on a graph. However, this solution might not be
optimal for the following reasons. First, the dispersion of node energies, $E_i$, leads
to the emergence of local minima where the population can be trapped
at low temperatures. From the perspective of diagonalization of rate matrix $\hat{W}$, it results
in a very broad spectrum of eigenvalues and, therefore, coexistence of different timescales.
The relaxation \emph{within} local minima occurs rapidly, but the
population transfer \emph{between} local minima is slow due
to the necessity to overcome potential barriers. This is illustrated
in Fig.~\ref{fig:graph}(b), where relaxation within local minima
(1-2-3 and 5-6-7) can be fast, but it is hard to overcome the potential
barrier (3-4-5) at low temperatures, which results in a very slow inter-minimum transfer. The natural criterion for the separation of timescales
for these two processes can be introduced as $\gamma=\delta E/T\gg1$,
where $\delta E=\left[N^{-1}\sum_i \left( E_i-\bar{E}\right)^2\right]^{1/2}$ is the standard deviation of node energies, and $\bar{E}=N^{-1}\sum_i E_i$ is their mean.
At $T\ll\delta E$ the eigenvalues of $\hat{W}$ can be spread over
many orders of magnitude, rendering the numerical diagonalization of the
rate matrix inaccurate due to numerical round-off errors.

The second reason is that the existence of a small parameter, $1/\gamma$,
might simplify the solution and analysis of the problem, and also might allow one to treat
much larger systems than those dealt with by the brute-force direct numerical diagonalization. Similarly to how the introduction
of a small parameter simplifies a problem in quantum mechanics by
allowing for a perturbative solution, we intend to explicitly use
the smallness of $1/\gamma$ to simplify the solution of the problem
and make it more physically transparent. 

\begin{figure*}
\includegraphics[width=4.7in]{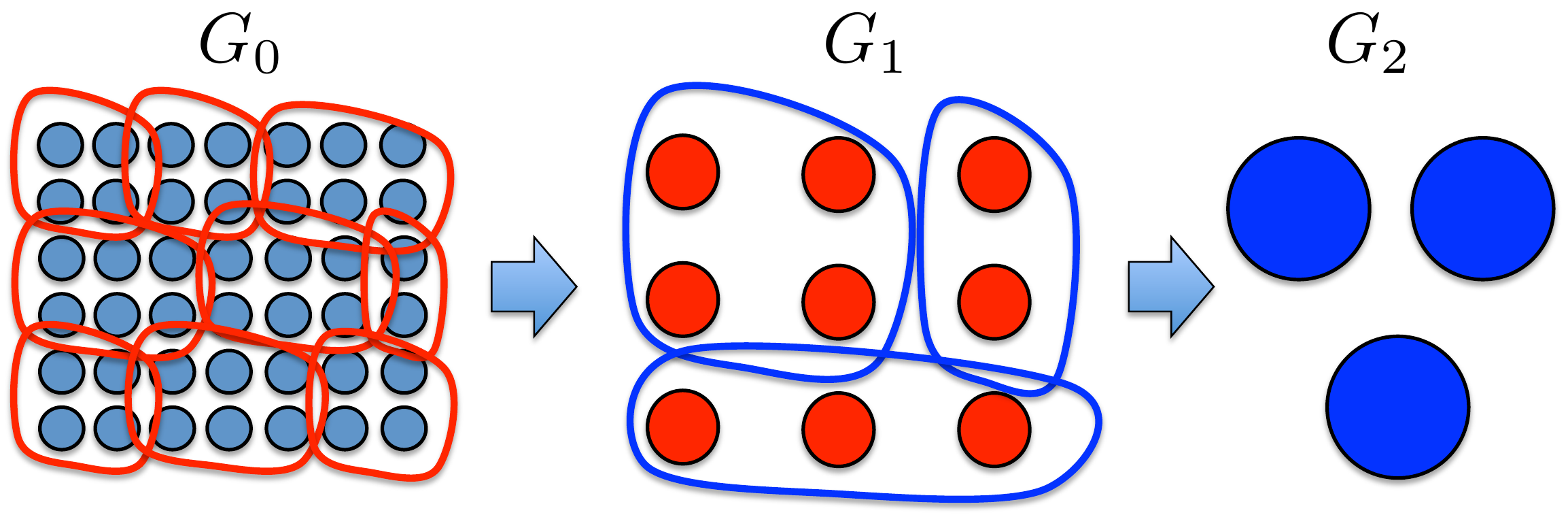}
\caption{\label{fig:hierarchy}
Schematic representation of consecutive RG transformations
leading to a hierarchy of coarse-graining levels. Closed contours show
how nodes are combined into basins.}
\end{figure*}

A straightforward approach is to explicitly separate the time scales
by first solving for population relaxation dynamics within each
local minima. Then, the inter-minima transfer problem is solved considering
population in each minima to be in its own quasi-equilibrium. This
is similar to transition state theory, where the timescale separation
allows one to consider both reactants and products equilibrated within
respective potential wells.\cite{Hanggi1990-251} Using the analogy
between energy dispersion of the graph and rain/water collection in
a rugged landscape we will refer to local minima as \emph{drainage
basins} or simply basins. Thus, we propose to divide the entire graph into
a collection of basins, and first solve for the population dynamics
\emph{within} each basin independently. Then, the population is considered
to be equilibrated within each basin and dynamics of population transfer
\emph{between} basins will be solved for. To make a step further, one can
notice that the population transfer between the equilibrated basins
can be again treated as a Markovian hopping process on a new graph
which describes \emph{inter-basin} hopping dynamics. This new graph
can be thought of as a result of coarse-graining of the initial graph.
The construction of this new graph out of the initial one is essentially an RG transformation (or step) applied to the original graph. This
RG step can be performed multiple times consecutively giving
rise to a \emph{hierarchy} of levels of coarse-graining. At each
such level, the population is being transferred between nodes, and each of these nodes represents an``equilibrated" basin of the graph belonging to the previous
hierarchy level. Fig.~\ref{fig:hierarchy} schematically depicts
the consecutive RG transformations applied to the original (microscopic) graph.
%============================================================ 
%============================================================ 
In this figure, the microscopic graph is denoted by $G_{0}$. The basins of this graph are outlined by the red contours, and upon
the RG step, each basin within $G_{0}$ becomes just a single node
(red circle) in a new graph, $G_{1}$. Once $G_1$ is constructed, the RG transformation can be applied to this new graph, yielding $G_{2}$ and so on. In what follows,
each graph within the RG hierarchy will be designated by $G_{h}$,
where index $h$ stands for the number of RG transformations applied consecutively
to the original graph $G_{0}$. The ordered set (lowest to highest)
of the exact eigenvalues of graph $G_{h}$ (e.g., obtained via numerical
diagonalization) is denoted by $\Lambda_{h}=\left\{ \lambda_{1},\lambda_{2},...,\lambda_{N}\right\} $. The set of eigenvalues corresponding to relaxation within basins of
$G_{h}$ is denoted by $\Lambda'_{h}$. Thus, the number of elements in $\Lambda'_h$ and $\Lambda_h$ is different by the number of basins in $G_h$. Further, we introduce a notation for the entire spectrum of the system, obtained by $H$ successive RG transformations, as
\begin{equation}
S_{H}=\bigcup_{h=0}^{H-1}\Lambda'_{h}\cup\Lambda_{H}.
\end{equation}
If RG transformations were exact, the following equality would hold 
\begin{equation}
S_{0}=S_{1}=S_{2}=...
\end{equation}
Since an explicit timescale separation is approximate, and therefore also the RG transformation, the validity of this expression depends on the magnitude of $\gamma$ at \emph{each} level of the hierarchy. In the remainder of the section we discuss the specific procedure of coarse-graining.

\subsection{Single RG step}

We propose the following algorithm for the coarse-graining procedure: 
\begin{enumerate}
\item In the entire graph, nodes are identified which are linked only to nodes
of higher energy. Each such low-energy node will be called the ``seed''
of a basin. 
\item For any other node of index $i$, a node $j$ is found so that the
following requirements are satisfied: (i) there exists an edge linking
nodes $i$ and $j$, (ii) $E_{i}>E_{j}$, and (iii) of all indices
$j$ satisfying the first two requirements we choose the one maximizing
$k_{ji}$. If node $j$ has already been associated with any basin (e.g.,
node $j$ is a seed), we associate node $i$ with the same basin. 
\item Repeat Step 2 until all the the nodes of the graph are associated
with basins. 
\end{enumerate}
Step 1 finds all the seeds and associate them with basins yet to be ``grown". The first execution of Step 2 associates with basins only the
nodes connected directly to basin seeds. Then, nodes connected to
seeds through a single node (i.e., via two edges) are associated
with basins and so on. At a very low temperature (i.e., $\gamma\gg1$), this
approach guarantees that a particle (e.g., an exciton), initially residing at node $i$, will preferentially relax to the seed of the basin node $i$ belongs to. Here, ``preferential'' means that we construct the basins using the maximum $k_{ij}$ requirement, neglecting all smaller $k_{ij}$. This can lead to a non-negligible probability of particle relaxation to other seeds. This is illustrated by initially putting an exciton on node $4$ of
the graph shown in Fig.~\ref{fig:graph}(b). As is seen, it can relax
to both seeds $2$ and $6$, but we associate node $4$ with the left
basin if $k_{34}>k_{54}$, and with the right basin if $k_{34}<k_{54}$. 

Once basins are formed, we construct a rate matrix $\hat{W}^{b}$ for each basin ($b$ is the basin index) as if it is isolated from the rest of the graph. In other words, $k_{ij}$ is set to zero if nodes $i$ and $j$ belong to different basins. Then, a typically small rate matrix for each isolated basin is numerically diagonalized
\begin{equation}
\hat{W}^{b}|n,b\rangle=-\tilde{\lambda}_{n}^{b}|n,b\rangle,\label{eq:BasinDiag}
\end{equation}
where the lowest eigenvalue vanishes exactly ($\lambda_{0}^{b}=0$)
because of the detailed balance, identically to the case of the entire
graph (see above). The eigenvectors, corresponding to these quasi-equilibrium
eigenmodes (one per basin), read as (right and left eigenvector, respectively) 
\begin{align}
(i|0,b\rangle &=Z^{-1}_be^{-\beta E_{i}}\delta_{i\in b},\nonumber\\
\langle0,b|i) &=\delta_{i\in b},
\end{align}
where $\delta_{i\in b}$ equals $1$ if node $i$ belongs to basin
$b$ and zero otherwise. The partition function of basin $b$ is given
by $Z_{b}=\sum_{i\in b}e^{-\beta E_{i}}$. The left and right eigenvectors are obviously orthonormal, i.e., $\langle0,b'|0,b\rangle=\delta_{bb'}$.

The problem of ``preferential'' relaxation can lead
to a significant underestimation of eigenvalues corresponding to
intra-basin relaxation, $\tilde{\lambda}_{n}^{b}$ ($n\neq0$). Indeed,
at very low temperatures the population of node 4 in Fig.~\ref{fig:graph}(b)
would decay with the rate constant of $k_{34}+k_{54}$. Associating node $4$ with
a specific basin (left or right), as it is done in the coarse-graining
procedure described above, would allow it to decay with rate $max(k_{34},k_{54})$
which can introduce a significant error up to a factor of $C$ - graph
connectivity. Indeed, for the specific example discussed here (one-dimensional graph, $C=2$), one has
\begin{equation}
max(k_{34},k_{54})\le k_{34}+k_{54}\le 2 \times max(k_{34},k_{54}),\label{eq:pref_ineq}
\end{equation}
so if $k_{34}$ and $k_{54}$ are not too different,  Eq.~(\ref{eq:BasinDiag}) produces significantly underestimated intra-basin eigenvalues. To correct for this, we recalculate eigenvalues of intra-basin
relaxation modes as
\begin{equation}
\lambda_{n}^{b}=-\langle n,b|\hat{W}|n,b\rangle.\label{eq:BasinDiag_corr}
\end{equation}
Using the language of perturbation theory in quantum mechanics, one
can say that these new eigenvalues are first-order corrected relative to
the zeroth-order approximation given by Eq.~(\ref{eq:BasinDiag}).
Eigenvalues $\lambda_{n}^{b}$ ($n\neq0$) constitute $\Lambda'_{h}$
at a given hierarchy level $h$, and are not modified any more by
any further (consecutive) RG transformations.

The rate matrix for the the coarse-grained graph, describing the population
transfer between basins, is constructed as follows. The rate constants
for population transfer between equilibrated basins are defined as
\begin{equation}
k_{b'b}=\langle0,b'|\hat{W}|0,b\rangle=\sum_{i\in b'}\sum_{j\in b}k_{ij}Z^{-1}_b e^{-\beta E_{j}}\label{eq:brates}
\end{equation}
Physically, this means that we take equilibrium populations of basin
$b$ and multiply them by rates of population transfer to basin $b'$.
It is easy to see that 
\begin{equation}
k_{b'b}/k_{bb'}=e^{-\beta(F_{b'}-F_{b})},
\end{equation}
where the Helmholtz free energy of a basin is defined as $F_{b}=-T\ln Z_{b}$. The new graph is the result of the RG transformation applied to the
graph of the previous hierarchy level. It is constructed as: 
\begin{enumerate}
\item Each node is the basin of the previous graph. 
\item The energy of each node is the free energy of the equilibrated basin of the previous graph. 
\item Rate constants of population transfer between nodes are given by Eq.~(\ref{eq:brates}) 
\item Initial populations of each node of the new graph are given by $\langle0,b|p_{0}\rangle$, where $|p_0\rangle$ is the vector of initial populations of the previous graph. 
\item Time-dependent population of each node is interpreted as the total
population of a corresponding equilibrated basin of the previous graph.
\end{enumerate}

\subsection{Consecutive RG steps as approximate diagonalization without dimensionality reduction}

In contrast to standard RG procedures in statistical physics, where
upon an RG step the microscopic information is typically lost,
we can keep the entire spectrum of eigenvalues during the coarse-graining procedure.
In the coarse-graining procedure we proposed above, the rate matrix $\hat{W}$
is partially diagonalized at each RG step in a sense that the first
RG step applied to $\hat{W}$ transforms the rate matrix to a block-diagonal
one. One of these blocks is exactly diagonal within, as it consists
of eigenvalues $\lambda_{i}^{b}$ ($i\neq0$), corresponding to the
intra-basin relaxation. The other block, non-diagonal within, is the
rate matrix for inter-basin transitions, constructed out of rates
defined by Eq.~(\ref{eq:brates}). Each consecutive RG step further decreases the size of this internally non-diagonal block and correspondingly increases
the size of the internally diagonal one keeping the overall dimensionality of the full rate matrix the same.
In other words, the RG approach introduced here is a method
to approximately diagonalize a certain class of matrices such as rate matrix
$\hat{W}$ at low temperature. It is, therefore, clear that Eq.~(\ref{eq:meq_eigen})
is still valid as long as the RG-based diagonalization of $\hat{W}$
is accurate. For example, Eq.~(\ref{eq:en_av}) can still be used
to find the time-resolved average energy of the system.

Clearly, only a finite number of consecutive RG transformations can be
applied to a finite system, since the size of the graph is rapidly decreasing with each consecutive level of the hierarchy (see below).
In fact, after a certain number of successive RG steps the direct diagonalization
is typically possible even for a very large initial system. The validity
and accuracy of the coarse-graining procedure can impose an additional
constraint on the maximum number of RG steps. Indeed, $\delta E$ - energy dispersion of graph nodes - can be defined at each level of hierarchy, resulting in the whole set of perturbation parameters, $\{1/\gamma_0,1/\gamma_1,...\}$. Even if $1/\gamma_0$ is sufficiently small to guarantee the validity of coarse-graining of the original graph, RG might become inaccurate at hierarchy level $h$ when $1/\gamma_h$ is on the order of $1$. We discuss this issue in Sec.~\ref{sec:flow}.

\section{Numerical results for coarse-graining\label{sec:num}}

To demonstrate the validity of the proposed RG approach at low temperatures,
we consider the following generic systems with energy disorder. Regular lattices, either one-dimensional chain (1D) or two-dimensional square (2D), are constructed. Only the nearest neighbors are linked
and periodic boundary conditions imposed, forming graph $G_{0}$
with connectivity $C=2$ and $C=4$ in the 1D and 2D cases, respectively.
Hopping rate constants are assigned to each edge of the graph by using
the modified version of the Miller-Abrahams (MA) model \cite{Miller1960-745}
\begin{equation}
k_{ij}=\begin{cases}
k_{ij}^{0}\exp(-\beta\Delta E_{ij}), & \Delta E_{ij}>0\,({\rm {\rm up-hop}}),\\
k_{ij}^{0}, & \Delta E_{ij}<0\,({\rm {\rm down-hop}}).
\end{cases}\label{eq:modMA}
\end{equation}
Node energies, $E_i$ ($i=1,...,N$), are assigned via sampling the Gaussian distribution with $\delta E=\langle (E-\bar{E})^{2}\rangle=1$. The mean node energy, $\bar{E}=N^{-1} \sum_i E_i$, is irrelevant for hopping dynamics since only relative node energies enter the expression for the rate constants.

In Eq.~(\ref{eq:modMA}), a single pre-exponential factor is assigned to each edge
\begin{equation}
k_{ij}^{0}=k_{ji}^{0}=\exp(-\beta\epsilon_{ij}^{a}),\label{eq:modMA_preexp}
\end{equation}
where $\epsilon_{ij}^{a}$ is an activation energy of the hopping
along edge $(i,j)$. Activation energies are assigned via sampling the standard uniform
distribution, $\epsilon_{ij}^{a}\rightarrow\mathcal{U}(0,1)$, so that $\epsilon_{ij}^{a}\in(0,1)$.
The presence of this temperature-dependent pre-exponential factor makes Eq.~(\ref{eq:modMA})
different from the classical MA model, the latter is recovered by setting 
all the activation energies to some fixed value, e.g., $\epsilon^a_{ij}=0$. The nature and
implications of this modification are discussed in the end of Sec.~\ref{sec:flow}. In the rest
of the section numerical results are presented for 1D and 2D systems
with the rate constants specified by Eqs.~(\ref{eq:modMA}) and (\ref{eq:modMA_preexp}).

\subsection{1D lattice}

\begin{figure*}
\includegraphics[width=6in]{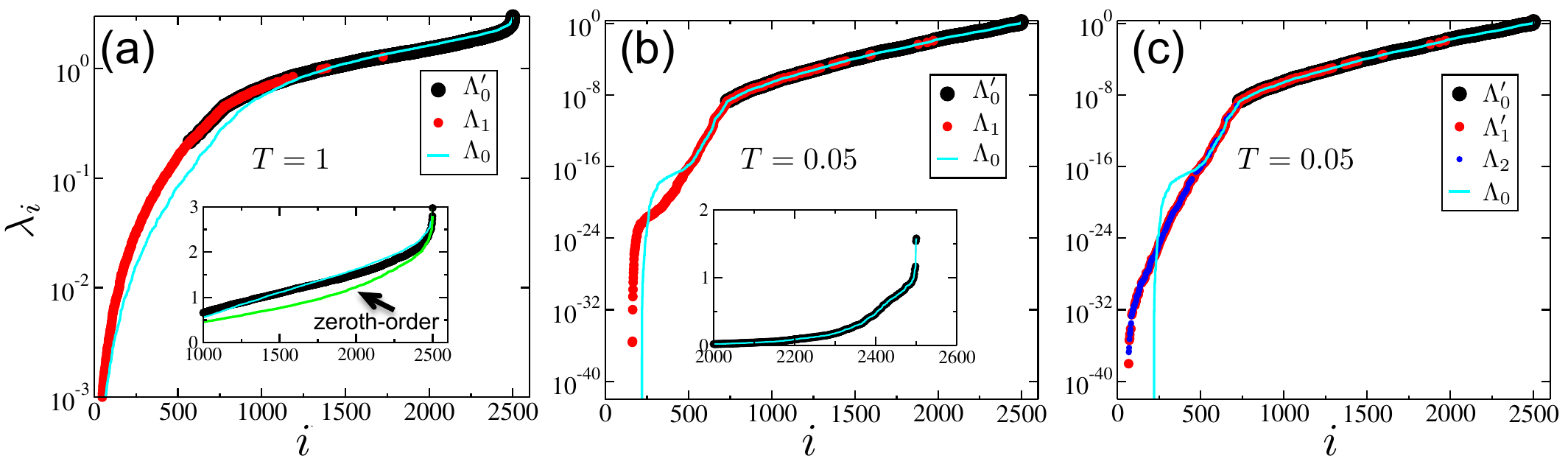}
\caption{\label{fig:1D_RG}
Comparison of the RG approach with direct diagonalization for the 1D chain, $\Lambda_{0}$. (a) Direct diagonalization versus a single RG step
at high temperature ($T=1$). (b) Direct diagonalization versus a single RG step at low
temperature ($T=0.05)$. (c) Direct diagonalization versus two RG steps at low temperature
($T=0.05$). Insets represent the same data as main graphs, but plotted in linear axes to emphasize the region of large eigenvalues. The inset in panel (a) also shows the eigenspectrum without the first-order perturbative correction of eigenvalues, obtained from Eq.~(\ref{eq:BasinDiag}) instead of Eq.~(\ref{eq:BasinDiag_corr}). }
\end{figure*}

Fig.~\ref{fig:1D_RG} presents the results of the RG-based calculations
of the eigenvalue spectrum for a 1D system with 2500 nodes. 
%============================================================ 
%============================================================ 
Panel (a) shows the spectrum obtained by the direct diagonalization
($\Lambda_{0}$), as well as the RG approach with a coarse-graining ($\Lambda_{0}'$,
$\Lambda_{1}$) at a temperature equal to the magnitude of the energy
disorder, i.e., $T=\delta E=1$. As expected, at $T/\delta E\gtrsim1$
the performance of the RG approach is not perfect. Specifically, it underestimates
the large eigenvalues originating with intra-basin relaxation (compare $\Lambda_0$ and $\Lambda'_0$ in the inset at $i\sim 2300$), because an exciton can escape a basin on the same timescale as the intra-basin relaxation occurs, which increases eigenvalues of ``intra-basin" modes. That RG neglects basin escape processes at
short times ($\Lambda_{0}'$) leads, in turn, to their effective emergence at long times
($\Lambda_{1}$), resulting in overestimated values for smaller eigenvalues (compare $\lambda_i$ at $i\sim 500$). In other words, the incomplete timescale separation at high temperatures can be thought of as an interaction between intra-basin relaxation and basin escape processes, which is not accounted for within RG. Treating this interaction within the second-order perturbation theory would shift high eigenvalues even higher, and low eigenvalues even lower, similarly to the standard second-order perturbative corrections to energy in quantum mechanics. Accordingly, RG underestimates high (intra-basin) and overestimates low (inter-basin) eigenvalues due to the lack of second-order perturbative corrections.

Numerical results for the significantly smaller temperature, $T=0.05$,
are shown in panels (b) and (c). As is seen, the performance of the RG
method improved, as compared to the large temperature case,
as RG results nearly coincide with the direct diagonalization results for
a large (higher) part of the spectrum [see also inset in panel (b)]. In fact, at such low temperatures
the RG approach outperforms the direct diagonalization. The kink seen
in the direct diagonalization spectrum at $\lambda_{i}\sim10^{-18}$, followed by a rapid
decrease, is due to numerical round-off errors: $\approx$220 eigenvalues
turned out to be negative in $\Lambda_{0}$. In comparison, a single
coarse-graining gives only $\approx$150 negative eigenvalues in $\Lambda_{1}$
[see panel (b)], which is seen by a lower position of the kink.
Furthermore, two consecutive RG steps applied to $G_0$ result in only $\approx$50 negative
eigenvalues in $\Lambda_{2}$, as is seen in panel (c) by the apparent absence
of the kink and very low onset of the rapid decrease in the RG eigenspectrum. 

The inset of Fig.~\ref{fig:1D_RG}(a) also shows the comparison of the first-order corrected eigenspectrum ($\Lambda'_0$), defined by Eq.~(\ref{eq:BasinDiag_corr}), and the zeroth-order eigenspectrum (green line), defined by Eq.~(\ref{eq:BasinDiag}). As was anticipated [see the discussion between Eq.~(\ref{eq:BasinDiag}) and Eq.~(\ref{eq:BasinDiag_corr})], the first-order corrected spectrum agrees better with the exact result, $\Lambda_0$, than the zeroth-order eigenspectrum. However, at lower temperatures the zeroth-order and first-order corrected spectra practically coincide (not shown). This is due to the fact that since the pre-exponential factors of the hopping rate constants are distributed according to Eq.~(\ref{eq:modMA_preexp}), at low temperatures any two given rate constants are almost always very different. This effectively reduces Eq.~(\ref{eq:pref_ineq}) to $max(k_{34},k_{54})\approx k_{34}+k_{54}$, and, therefore, even the zeroth-order approximation works well. In what follows, we use the first-order corrected eigenvalues even though the zeroth-order approximation is quite accurate at low temperatures.

\subsection{2D lattice}

\begin{figure*}
\includegraphics[width=6in]{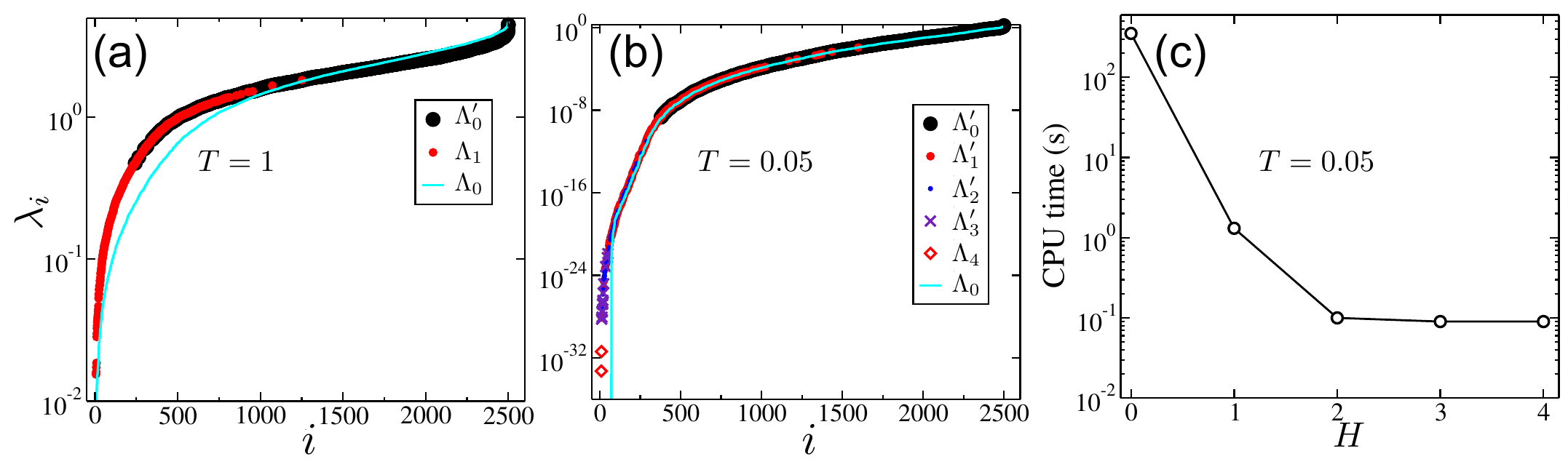}
\caption{\label{fig:2D_RG}
Comparison of the RG approach and direct diagonalization ($\Lambda_0$) for the 2D
square lattice of size $50\times50=2500$. (a) Direct diagonalization versus a single
RG step at $T=1$. (b) Direct diagonalization versus four RG steps at $T=0.05$. (c) CPU
time versus number of RG steps for $T=0.05$.}
\end{figure*}

Numerical results for the 2D square lattice with $50\times50=2500$
nodes are shown in Fig.~\ref{fig:2D_RG}.
%============================================================ 
%============================================================ 
Similarly to the 1D case (Fig.~\ref{fig:1D_RG}), the RG approach
performs poorly when $\gamma$ is not large, as is shown in panel
(a). Panel (b) demonstrates the much better RG performance at a lower temperature,
where the RG transformation has been applied four consecutive times.
Numerical round-off errors are again seen as a vertical drop in the lower part of the direct diagonalization spectrum.

One can notice that at the same low temperature and number of nodes in the lattice,
the 1D system [Fig.~\ref{fig:1D_RG}(c)] possesses a broader spectrum than the 2D one [Fig~\ref{fig:2D_RG}(b)]. The reason for this is that 1D diffusion is an exact lower boundary
for $n$-dimensional diffusion, since hopping between any two distant nodes in $n$ dimensions can always
be considered as hopping in 1D but along the ``shortest'' path, i.e.,
the path with highest rate constants.\cite{Bouchaud1990-127} 

The necessary condition for the coarse-graining procedure developed here to be valid is to have a clear separation of time-scales between intra- and inter-basin dynamics. Figs.~\ref{fig:1D_RG} and \ref{fig:2D_RG} present a contradiction to this, since even at low temperature, where the RG procedure is seen to perform well, the eigenvalues from different coarse-graining levels [e.g., $\Lambda'_0$, $\Lambda'_1$, $\Lambda'_2$ and $\Lambda'_3$ in Fig.~\ref{fig:2D_RG}(b)] are seen to significantly overlap in magnitude. In fact, this observation does not contradict the previously mentioned condition, since the time-scale separation is only required for modes localized within the same small portion of the graph. An intra-basin relaxation mode within one portion of the large graph does not interact with an inter-basin hopping in some other distant portion of the graph. Therefore, the time-scale separation for these modes is not required for the RG procedure to work. Again, using the language of perturbation theory in quantum mechanics, one can say that for the hybridization of two interacting states to be weak, either the energy separation between the two states has to be large (i.e., large time-scale separation), or the interaction has to be weak. The latter condition resolves the apparent contradiction with overlapping time-scales in Figs.~\ref{fig:1D_RG} and \ref{fig:2D_RG}, since overlapping eigenvalues belong to intra- and inter-basin modes localized in different portions of the graph. 

\subsection{Computational performance and scaling}

As described above in detail, the RG procedure consists of splitting the entire graph into small basins, and then finding the intra-basin relaxation modes within each small basin. Splitting of the graph into basins requires a few consecutive scans of the entire graph, and the number of these scans is on the order of the average size of a basin, $n_h$, where $h$ specifies the current level of coarse-graining hierarchy. Thus, the computer time needed for the splitting of the graph can be estimated as $\sim n_h\times N_h$, where $N_h$ is the number of sites in the entire graph of hierarchy level $h$. Direct ``brute-force" diagonalization of a matrix of size $m$ costs on the order of $m^3$, and since the number of basins is $\sim N_h/n_h$, the computational cost of diagonalization of all the basins within the graph is $\sim n^3_h\times N_h/n_h=n^2_h\times N_h$. Some additional minor computational overhead is present, but its cost is also growing linearly with the size of the graph, $N_h$. Since the average basin size does not depend on the size of the graph, the overall computational cost of the RG step is
\begin{equation}
t_h\sim N_h,~~h=0,1,...,H-1,
\end{equation}
where $N_h$ can be written as
\begin{equation}
N_h=\frac{N}{\prod^{h-1}_{h'=0} n_{h'}}\approx \frac{N}{n^h_0},
\end{equation}
where $N\equiv N_0$ is the size of the initial microscopic graph.
The last approximate equality is obtained assuming that the average size of basins is not too different at all the levels of hierarchy, so we take $n_h\approx n_0$. Summing up all the computational times for all the RG steps yields
\begin{equation}
t_{RG}=f(H)N,\label{eq:RG_cost}
\end{equation}
where
\begin{equation}
f(H)\sim\sum^{H-1}_{h=0}1/n^h_0=\frac{n_0-n^{1-H}_0}{n_0-1}.
\end{equation}
Thus, $f(H)$ is a slowly growing function of $H$ which rapidly converges to a constant value of $f(H\rightarrow\infty)\sim n_0/(n_0-1)$.
 
The direct diagonalization of the entire graph at the highest level of the hierarchy costs as
\begin{equation}
t_H\sim N^3_H=\left(\frac{N}{\prod_{h=0}^{H-1} n_h}\right)^3\approx \left(\frac{N}{n^H_0}\right)^3.\label{eq:DD_cost} 
\end{equation}
It is readily seen by comparing Eqs.~(\ref{eq:RG_cost}) and (\ref{eq:DD_cost}) that if the original microscopic graph is large, then $t_H\gg t_{RG}$, since $t_H$ grows cubically with $N$. The overall computational time is thus dominated by $t_H$, and it decays exponentially with the number of hierarchy levels. However, since $t_{RG}$ grows and $t_H$ decays with $H$, it is expected that at a certain level of hierarchy, dependent on the size of the original graph, computational time becomes dominated by $t_{RG}$, which depends on $H$ only weakly.   

Fig.~\ref{fig:2D_RG}(c) shows the overall computational time required to
obtain the entire spectrum of hopping dynamics with $H=0,1,2,3$ or $4$ consecutive RG steps.
The hierarchical RG procedure has been implemented into a Fortran 90 code,
and all the calculations have been performed using a single core of an Intel(R) Xeon(R) CPU
(2.66GHz). As is expected from the scaling considerations above, the direct diagonalization (i.e., $H=0$) is not only plagued by the numerical round-off errors at low temperatures, but is also very slow, since a single very large matrix has to be diagonalized. Each consecutive RG step reduces the computational cost exponentially. As predicted above, at sufficiently large hierarchy level the computational cost saturates, e.g., at $H=2$ for $N=2500$ in Fig.~\ref{fig:2D_RG}(c). It is also worth noting that since diagonalization of many small matrices (intra-basin relaxation) within each RG step is absolutely independent, the efficient parallelization of the code is possible, which can result in significant lowering of function $f(H)$ in Eq.~(\ref{eq:RG_cost}) (not implemented in this work).

\subsection{Time-resolved luminescence}

To illustrate how the developed RG approach can be used to evaluate
spectroscopic observables, we simulate the time-resolved photoluminescence experiment for a system where an exciton
migrates within an energy-disordered lattice. Specifically, we evaluate
the time-resolved ensemble-average of photoluminescence energy. If an exciton is localized
at node $i$ of energy $E_{i}$, it can recombine radiatively emitting
a photon of energy $\hbar\omega=E_{i}$. Therefore, if we assume
for simplicity that the rate of radiative recombination and its quantum yield do not depend
on exciton energy, then the ensemble-average photoluminescence energy at time $t$
is given by $\langle E(t)\rangle=\langle E|p(t)\rangle$,
i.e., by Eq.~(\ref{eq:en_av}) for the average energy of an exciton.

\begin{figure*}
\includegraphics[width=4.5in]{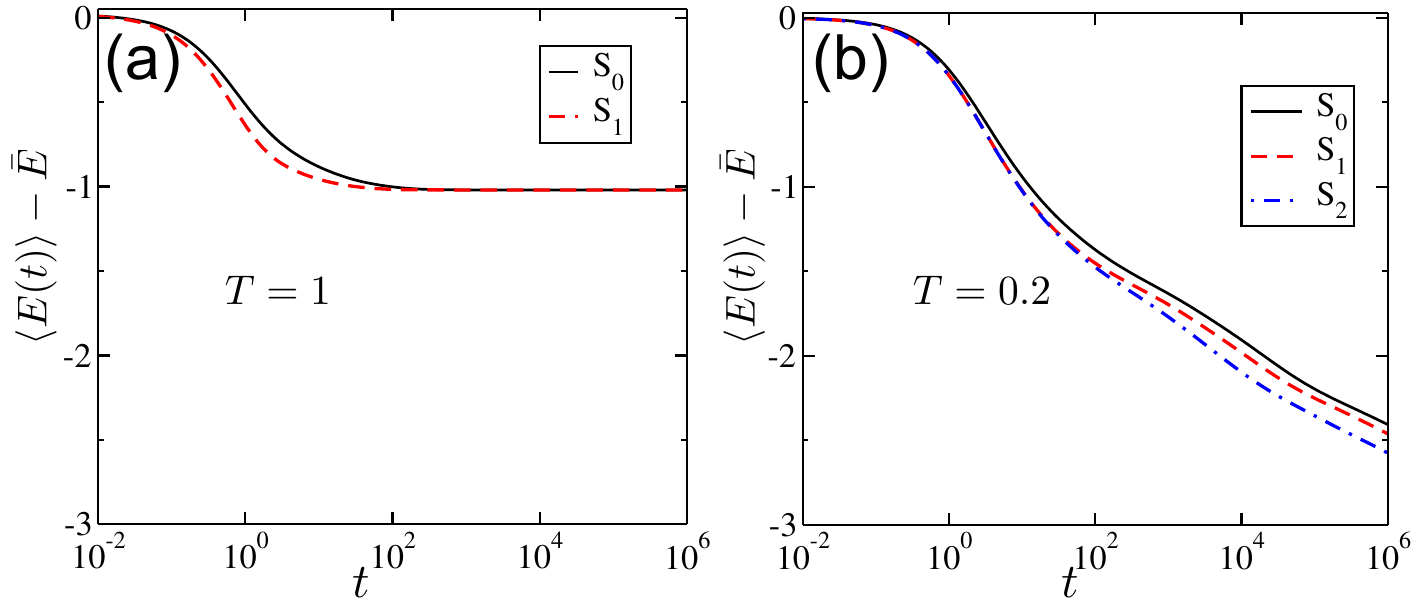} \caption{\label{fig:eav} Time-resolved ensemble-averaged energy of an exciton for the 2D graph of $50^{2}=2500$ nodes. Temperature is set to $T=1$ (a) and $T=0.2$ (b).}
\end{figure*}

The time-dependent ensemble-average energy of excitons within a 2D system of $50\times 50=2500$ nodes, measured relative to $\bar{E}$, is shown in Fig.~\ref{fig:eav}.
%============================================================ 
%============================================================ 
For definiteness, all the nodes are assumed to be populated with
equal probability at $t=0$, i.e., $|p_0\rangle=N^{-1}\sum_i |1)$. This initial state results in $\langle E(t)\rangle-\bar{E}=0$ at $t=0$ for both the
exact solution obtained via direct diagonalization, $S_{0}$, as well
as approximate RG-based ones, $S_{h}$ ($h>0$). At short times, the
agreement between exact and approximate solutions is still good in
panel (a), but at longer times ($t\sim1-10$) the RG approach is inaccurate
as expected for this high-temperature case, $T=1$. Within the same
time domain, the agreement of the RG approach and the direct diagonalization
is better in panel (b), where temperature is significantly lower ($T=0.2$),
and, therefore, RG is expected to produce more accurate results. At large
times the system is equilibrated in panel (a), so RG and direct
diagonalization results coincide again. At lower temperatures, panel
(b), the equilibration of the system is taking significantly longer and
having not been reached at maximum observation time of $t=10^{6}$.  

\section{Renormalization group flow\label{sec:flow}}

\begin{figure}
\includegraphics[width=3in]{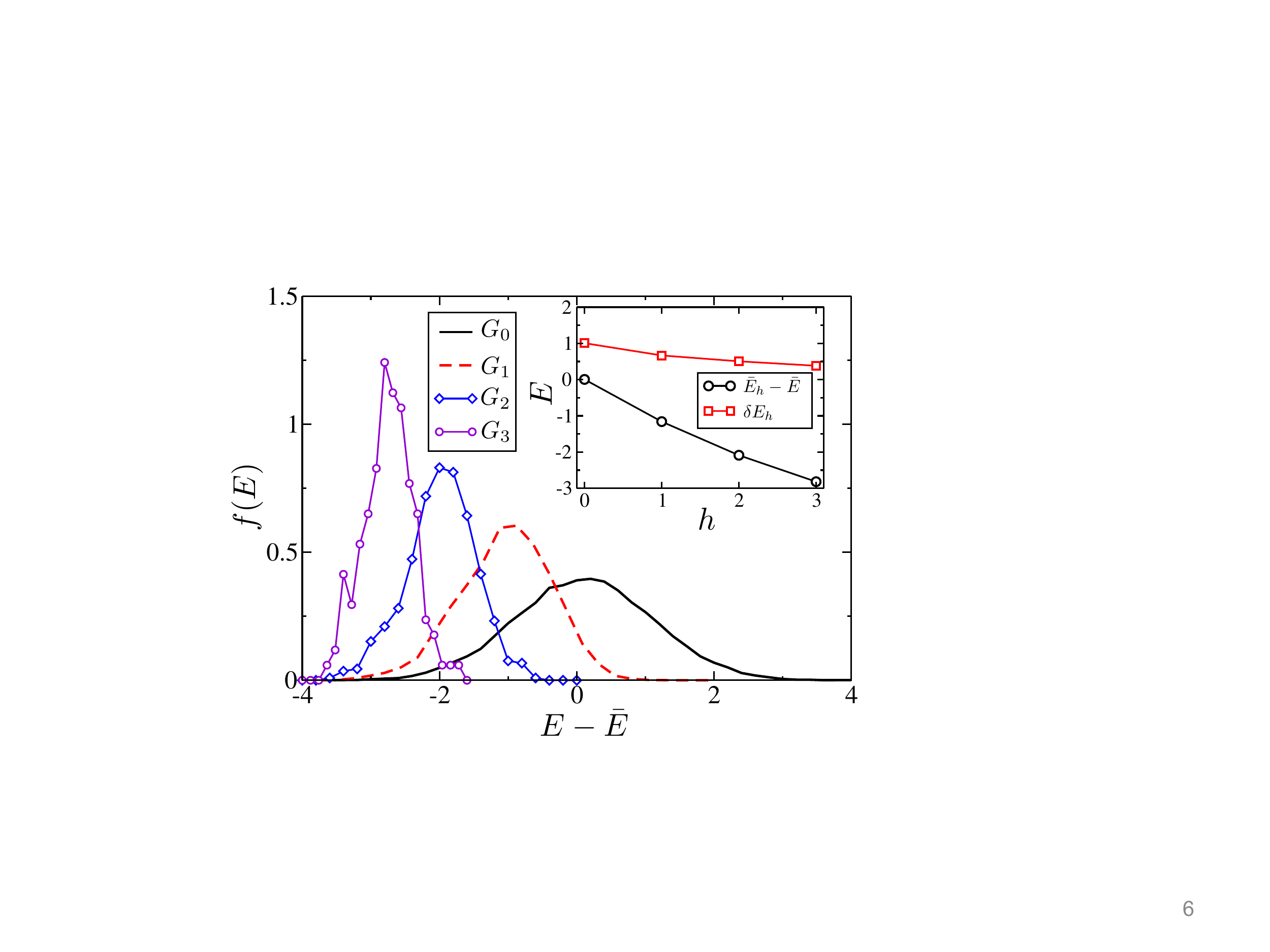}
\caption{\label{fig:Edistr}
Energy distribution of nodes corresponding to hierarchy
levels from $h=0$ (initial graph) to $h=3$ for the 2D system consisting
of $200\times200=40000$ nodes at $T=0.05$. The inset shows mean
(black circles) and the standard deviation (red squares) corresponding
to the distributions in the main figure. The size of the graphs for
the each hierarchy level are 40000, 8046, 1123 and 142 for $G_{0}$,
$G_{1}$, $G_{2}$ and $G_{3}$, respectively.}
\end{figure}

The analysis of dynamics of a system on different time- and length-scales,
as well as interaction between these scales, can often be conveniently
done by studying the RG flow -- the evolution of system parameters 
(e.g., energy dispersion, rate constants) under consecutive RG transformations.
A good example is the RG flow of the distribution of node energies. 
Fig.~\ref{fig:Edistr} shows the distribution of node energies for the original microscopic graph, $G_{0}$,
as well as the energy distributions for graphs obtained from $G_{0}$
by consecutive RG transformations: $G_{1}$, $G_{2}$ and $G_{3}$.
%============================================================  
%============================================================ 
All energies are measured relative to $\bar{E}$, so the distribution of node energies for $G_0$ is just a Gaussian distribution (by construction) centered around $E=\bar{E}$ and with $\delta E=1$.
Comparison of node energy distributions for $G_0$, $G_1$, $G_2$ and $G_3$ reveals the RG flow toward lower mean energies
and lesser standard deviations, i.e., the distribution grows sharper
under consecutive RG transformations. The extracted mean node energies and
standard deviations are plotted by black circles and red squares,
respectively, in the inset. The decrease of the mean node energy, $\bar{E}_h$, with $h$ can be
easily understood from the fact that at very low temperatures energies
of nodes of graph $G_h$ are essentially the energies of basin seeds of $G_{h-1}$, which are
found by requiring each such point to be energetically lower than
its immediate neighbors. This flow of $\bar{E}_h$ with $h$ is in fact reflected in Fig.~\ref{fig:eav} as the {\em spectral diffusion} -- the decay of the ensemble-average photoluminescence energy with time. Indeed, the higher levels of the RG hierarchy correspond to longer observation times and, therefore, the dependence of average exciton energy on $h$ and on $t$ are expected to be similar.

The reduction of the standard deviation of node energies upon
the RG step is less trivial and can be understood from the \emph{order
statistics}. Specifically, the mean and the standard deviation of
a minimum of $n$ random variables, each sampled independently from the standard normal distribution, is given by (in the limit of $n\rightarrow\infty$) \cite{Sterling2001}
\begin{align}
\bar{E} &=-\left(2\log n\right)^{1/2},\nonumber \\
\delta E &=\left(2\log n\right)^{-1/2},\label{eq:OrdStat}
\end{align}
As was mentioned above, at low temperatures, node energies of a graph obtained by the RG transformation are approximately equal to the energies of the seed nodes of the previous graph. In turn, each seed is found by identifying a node with minimum energy out of each immediate neighborhood. This procedure is very reminiscent of how a minimum value is found within each sample of size $n$ in the order statistics. Thus, we expect the mean node energy, $\bar{E}_h$, and the standard deviation, $\delta E_h$, to follow Eq.~(\ref{eq:OrdStat}) with $n$ approximately given by the average size of the basin. Eq.~(\ref{eq:OrdStat}) suggests the reduction
of both mean and standard deviation of node energies upon the RG transformation, which is indeed in qualitative agreement with the numerical results
(inset of Fig.~\ref{fig:Edistr}). Quantitatively, the average basin
size in the original graph is $n=\frac{N(G_{0})}{N(G_{1})}\approx5$,
so that Eq.~(\ref{eq:OrdStat}) yields $\delta E_1\approx0.56$ and
$\bar{E}_1\approx-1.8$. The numerical results (from the inset)
are $\delta E\approx0.67$ and $\bar{E}\approx-1.2$, so
the agreement is good considering the asymptotic ($n\rightarrow\infty$) nature of Eq.~(\ref{eq:OrdStat}).

\begin{figure}
\includegraphics[width=3in]{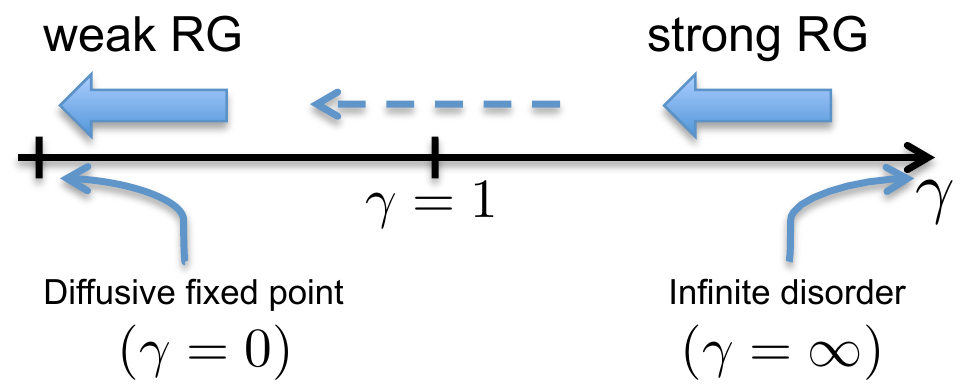}
\caption{\label{fig:RGflow}
Schematic representation of the RG flow of $\gamma=\delta E/T$ in strong and weak disorder limits.}
\end{figure}

Fig.~\ref{fig:RGflow} schematically depicts the RG flow of $\gamma=\delta E/T$,
i.e., how $\gamma$ changes when consecutive RG transformations are
applied.
%============================================================ 
%============================================================ 
The RG approach developed in this work is valid only at $\gamma\gg1$,
as demonstrated above, and, therefore, one can call it strong-disorder
RG. Fig.~\ref{fig:Edistr} and Eq.~(\ref{eq:OrdStat}) suggest that
upon consecutive RG transformations the value of $\gamma$ flows to lower magnitudes,
as illustrated in Fig.~\ref{fig:RGflow}. This imposes a natural
restriction on the maximum number of accurate RG transformations, since
above a certain number of transformations one has $\gamma\sim 1$,
and the strong-disorder RG becomes inaccurate.

On the other hand, the \emph{weak} disorder RG, discussed by Deem and Chandler,\cite{Deem1994-911}
demonstrates the same direction of the flow, i.e., reduction of $\gamma$ upon each consecutive RG step at $\gamma\ll 1$. In fact, the weak RG flow in this domain ultimately converges to the stable fixed point at $\gamma=0$, which is the pure diffusion. Neither of these RG approaches is valid in the
intermediate disorder regime, $\gamma\sim1$. However, the exact
solution for the 1D model suggests that the long-time behavior of
1D model at \emph{arbitrarily} strong gaussian disorder is always
purely diffusive.\cite{Bouchaud1990-127,Deem1994-911} Furthermore,
the long-time hopping dynamics in 1D model is an exact lower boundary for
models with higher dimensionality.\cite{Bouchaud1990-127} Therefore,
observing hopping in a model with gaussian energy disorder at increasing
length- or time-scale is expected to always correspond to a decreasing effective
disorder. We therefore put a dotted arrow in Fig.~\ref{fig:RGflow}
to reflect the expected flow direction at $\gamma\sim1$, even though
we are not aware of any RG approach valid in this region.

\begin{figure}
\includegraphics[width=2.8in]{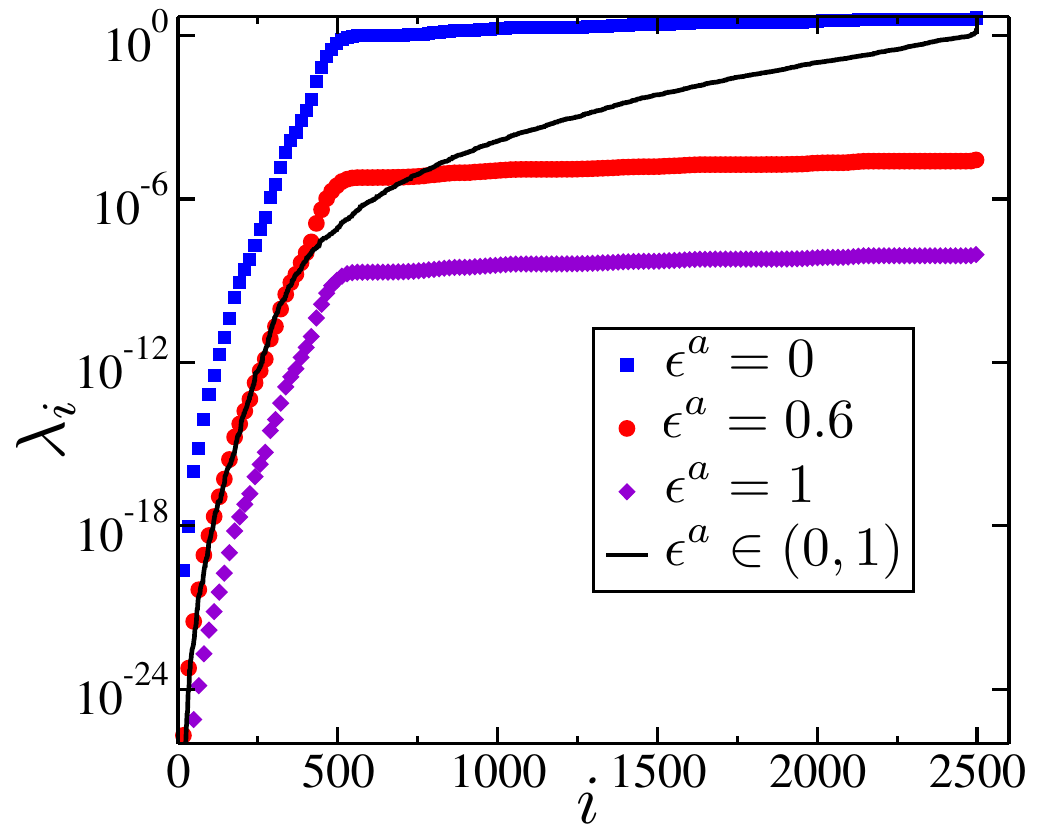}
\caption{\label{fig:MAvsMMA}
The eigenspectrum for the modified MA model ($\epsilon_{ij}^{a}\rightarrow\mathcal{U}(0,1)$,
black line), as well as for the classical MA model with the fixed pre-exponential factor
obtained from Eq.~(\ref{eq:modMA_preexp}) by fixing the activation
energy: $\epsilon_{ij}^{a}=0$ (blue squares), $\epsilon_{ij}^{a}=0.6$
(red circles) and $\epsilon_{ij}^{a}=1$ (magenta diamonds). All the
eigenspectra are evaluated for the 2D system with 2500 nodes at $T=0.05$
applying two consecutive RG steps. }
\end{figure}

Finally, we briefly discuss the RG flow of the distribution of the
pre-exponential factors of rate constants. Pre-exponential factors are all {\em degenerate} (i.e., the same) within the classical MA model. However, it is quite obvious that even
if this degeneracy is present in $G_0$, the pre-exponential factors of the rate constants of $G_1$ will have a certain finite dispersion
since pre-exponential factors of inter-basin rate constants are determined by
activation barriers of hopping between basins, and the heights of these activation
barriers are random variables themselves. Thus, it is expected that the degeneracy
of the pre-exponential factors (i.e., their non-randomness) is {\em irrelevant} in the RG sense, i.e., the system
flows away from this degeneracy upon consecutive RG transformations.
To illustrate this, we compare the spectra of eigenvalues for the modified and
classical MA models, the latter obtained from the modified one by setting
the activation energy to a certain fixed non-random value. Fig.~\ref{fig:MAvsMMA}
shows these spectra ($S_{2}$ at $T=0.05$ for the 2D lattice
with 2500 nodes) for the modified MA model (black line) with $\epsilon_{ij}^{a}$
sampled uniformly within $(0,1)$ interval, as well as 
for the classical MA model with the activation energy set to $\epsilon_{ij}^{a}=1$
(magenta diamonds), $\epsilon_{ij}^{a}=0.6$ (red circles) and $\epsilon_{ij}^{a}=0$
(blue squares). 
%============================================================ 
%============================================================ 
It is seen that in the high-$\lambda$ part of the spectrum, the
modified and classical MA models produce qualitatively different results.
Specifically, the classical MA model produces an almost flat spectrum
due to the high degree of degeneracy of eigenvalues. However, the
lower part of the spectrum is qualitatively the same for the two models, except for a
vertical shift. The origin of this shift is clear since, e.g., at $\epsilon_{ij}^{a}=1$
all the rate constants are lower than those for the activation energy 
uniformly distributed between $0$ and $1$. Therefore, the spectrum with
this fixed activation energy is expected to be generally lower than
the one with finite dispersion of activation energies (compare the
black line and magenta diamonds in Fig.~\ref{fig:MAvsMMA}). A similar
argument explains why the eigenspectrum with $\epsilon_{ij}^{a}=0$
is generally higher than the one with the dispersion of pre-exponential factors present (compare the
black line and blue squares in Fig.~\ref{fig:MAvsMMA}) In fact,
it is possible to numerically find the fixed activation energy, $\epsilon_{ij}^{a}=0.6$,
at which the lower portion of the eigenspectra of the standard and modified MA models practically coincide. Therefore, the degeneracy, or non-randomness, of the pre-exponential factors is \emph{irrelevant} in the RG sense, which is exactly the reason why we adopted the modified MA model from the onset.

\section{Conclusion\label{sec:conclusion}}

In this paper, we formulate and computationally implement a numerical
real-space RG approach to study the hopping dynamics on energy-disordered lattices
at low temperature. Our approach is similar to methods developed in Refs.~\onlinecite{Dasgupta1980-1305,Igloi2005-277,Lee2009-046210,Amir2010-070601},
which were originally introduced to treat disordered Ising-type models. This
similarity stems from the fact that in both cases fast degrees of
freedom can be treated independently from slow ones, giving rise
to renormalization of the latter. However, we emphasize that in this
work the fast degrees of freedom are not actually ``integrated out'',
since the entire spectrum is recovered, not just its low-energy portion.
The other difference of the approach, developed in this work, from that in Refs.~\onlinecite{Dasgupta1980-1305,Igloi2005-277,Lee2009-046210,Amir2010-070601} is that the former is \emph{dynamical}
in a sense that the dynamics in the form of eigenvalues and eigenvectors of relaxation modes
is recovered at all timescales, and not just equilibrium properties
of a system of interest.

In this work, the developed RG approach is applied to discrete systems,
represented by graphs. In fact, in the case of strong disorder the
approach can be straightforwardly applied to continuous systems. Indeed,
even if the microscopic system is initially continuous, the population
trapping in local minima leads to the appearance of the discrete-like
dynamics, i.e., hopping between potential minima of the continuous landscape. In other words,
an RG transformation can still be introduced, and its first application
to the continuous system in the regime of strong disorder immediately leads to hopping on a graph. Thus, the continuity of a system
is irrelevant in the RG sense in the case of strong disorder. Interestingly, it
is exactly the opposite in the weak disorder limit, where discreteness becomes irrelevant and continuity finally ``sets in'' leading to pure diffusion at large timescales.

We expect the developed approach to become useful in studying hopping
dynamics in various systems of practical interest including polymer
films and nanoparticle aggregates. A combination of this method with the weak-disorder RG group\cite{Bouchaud1990-127,Deem1994-911} might become especially fruitful. Indeed, as was discussed above, the validity range of strong-disorder RG is limited by the fact that the system``flows away" from strong disorder upon consecutive RG transformations. However, once the disorder is not strong, the weak-disorder approach can be applied (at least approximately), ultimately leading to the pure diffusion fixed point (see Fig.~\ref{fig:Edistr}). Therefore, the combination of the strong-disorder RG method developed in this work with the weak-disorder RG approach might yield a universal methodology capable of treating hopping dynamics in systems with an arbitrary strength of disorder. Furthermore, since the attractive RG fixed point -- pure diffusion -- possesses an elementary analytical solution, the universal methodology can be applied to systems of infinite extent, yielding the large-scale diffusion coefficient, determined by small-scale hopping rate constants.\cite{Bouchaud1990-127}  

K.A.V. is thankful to Y. Dubi, A. Zhugayevych and J. Bjorgaard for useful discussions
and comments on the manuscript. Los Alamos National Laboratory, an
affirmative action equal opportunity employer, is operated by Los
Alamos National Security, LLC, for the National Nuclear Security Administration
of the U.S. Department of Energy under contract DE-AC52-06NA25396.
V.Y.C acknowledges support by the National Science Foundation under Grant No. CHE-1111350.

%\bibliography{/Users/velizhan/Work/Bib/main}


\end{document}